\documentclass[article]{elsarticle}

\usepackage{lineno,hyperref}
\modulolinenumbers[5]

\journal{Annals of Physics}

\begin{document}

\begin{frontmatter}

\title{ Neutron rich matter in the laboratory and in the heavens after GW170817}

\author{C. J. Horowitz}
\ead{horowit@indiana.edu}
\address{Center for Exploration of Energy and Matter and Department of Physics, Indiana University, Bloomington, IN 47405, USA}




\begin{abstract}
The historic observations of the neutron star merger GW170817 advanced our understanding of r-process nucleosynthesis and the equation of state (EOS) of neutron rich matter.  Simple neutrino physics suggests that supernovae are not the site of the main r-process.  Instead, the very red color of the kilonova associated with GW170817 shows that neutron star (NS) mergers are an important r-process site.  We now need to measure the masses and beta decay half-lives of very neutron rich heavy nuclei so that we can more accurately predict the abundances of heavy elements that are produced.  This can be done with new radioactive beam accelerators such as the Facility for Rare Isotope Beams (FRIB).  GW170817 provided information on the deformability of NS and the equation of state of dense matter.  The PREX II experiment will measure the neutron skin of ${}^{208}$Pb and help constrain the low density EOS.   As the sensitivity of gravitational wave detectors improve, we expect to observe many more events.  We look forward to exciting advances and surprises!
\end{abstract}


\end{frontmatter}

%\linenumbers

\section{Introduction: neutron rich matter}
 Compress almost anything enough to densities of $10^{11}$ g/cm$^3$ or more and electrons react with protons to make neutron rich matter. This material is at the heart of many fundamental questions in nuclear physics and astrophysics.
 \begin{itemize}
\item
 What are the high density phases of QCD?
\item
Where did chemical elements come from?
\item
What is the structure of many compact and energetic objects in the heavens, and what determines their electromagnetic, neutrino, and gravitational-wave radiations?
\end{itemize}

Neutron rich matter is a very versatile material.  We are interested in it over a tremendous range of density and temperature were it can be a gas, liquid, solid, plasma, liquid crystal (nuclear pasta), superconductor ($T_c\approx 10^{10}$ K!), superfluid, color superconductor...
We can probe neutron rich matter observationally, in the laboratory, and computationally.  With multi-messenger astronomy: ``seeing'' the same event with very different probes can lead to fundamental advances. Often %X-ray photons come from near the {\it solid} neutron star crust, 
supernova neutrinos come from a low density warm {\it gas} of neutron rich matter, and gravitational waves come from energetic motions of the {\it liquid} interior of (merging) neutron stars.

Nuclei are liquid drops so most laboratory experiments probe liquid neutron rich matter.  We will discuss electroweak measurements of the neutron skin thickness of neutron rich nuclei.  In addition new radioactive beam accelerators such as the Facility for Rare Isotope Beams (FRIB) provide laboratory access to very neutron rich nuclei.  Finally, chiral effective field theory \cite{Chiral_review} allows computation of properties of low density neutron rich matter.  Here there are important contributions from possibly poorly constrained three neutron forces.  

However, the chiral expansion does not converge at high densities.  Furthermore there is an important sign problem that so far prevents accurate lattice QCD results at high densities.  This strongly limits our ability to calculate properties of dense neutron rich matter from first principles.  These fundamental computational limitations dramatically increase the importance of astronomical observations of dense neutron rich mater.

The extraordinary observations of the neutron star merger GW170817 \cite{LIGO1,LIGO2} have many implications. In Sec.~\ref{Sec2} we discuss  r-process nucleosynthesis.  Next in Sec.~\ref{Sec3} we consider the equation of state of neutron rich matter.  Finally, in Sec.~\ref{Sec4} we summarize our understanding of the composition and phases of cold dense QCD matter after GW170817.

\section{Nucleosynthesis}
\label{Sec2}
The optical counterpart or kilonova associated with the gravitational wave event GW170817 originally had a blue spectrum.  This suggests there was a component of the ejecta from the neutron star merger that was lanthanide free.   We expect material that is not very neutron rich, with an electron fraction $Y_e>0.25$, to not produce many lanthanides.  In contrast, neutron star material is very neutron rich with $Y_e\approx 0.1$ or lower.  Neutrinos emitted during the merger could have played an important role in raising the $Y_e$ of this ejecta via $\nu_e+n\rightarrow p + e$.  Note that antineutrinos can lower $Y_e$ via $\bar\nu_e+p\rightarrow n + e^+$.  However when the material starts very neutron rich, the rate of $\nu_e$ capture is expected to dominate over $\bar\nu_e$ capture.

Unfortunately, it may be very difficult to detect neutrinos from a neutron star merger.  The number of neutrinos radiated during a merger may be roughly comparable to the number radiated during a core collapse supernova (SN).  Indeed, about 20 events were detected from SN1987A.  However, NS mergers are rarer than SN and are likely to be much further away.  Indeed GW170817 was nearly a thousand times further away than SN1987A.  A factor of one thousand in distance corresponds to a count rate lower by a factor of one million.  Thus it is extremely difficult to detect NS merger neutrinos at the expected distances of tens of Mpc.

Therefore, it is very important to observe in detail neutrinos and antineutrinos from the next galactic SN.  This is important both to learn about neutrino oscillations \cite{SN_osc} and other neutrino physics in an astronomical environment and to learn about the role of neutrinos for nucleosynthesis in SN.   Furthermore what we learn from SN neutrinos may be directly relevant for neutrinos from NS mergers.  For example, we could discover an unexpected neutrino oscillation in SN that could impact nucleosynthesis in both SN and NS mergers.  

\subsection{Detecting SN neutrinos}
Core collapse SN radiate the gravitational binding energy of a NS (almost $0.2M_\odot c^2$) as $\approx 10^{58}$ neutrinos.  We detected about 20 events from the historic SN1987A \cite{PhysRevD.70.043006}.  Several thousand events are expected from the next galactic SN in Super-Kamiokande \cite{Scholberg2012}.  This is a large $H_2O$ detector that is good at measuring electron {\bf anti-neutrinos} $\bar\nu_e$ via inverse beta decay $\bar\nu_e+p\rightarrow n + e^+$.   Furthermore, the planned Hyper-Kamiokande detector will be a significantly larger version of Super-K and it could detect $\approx 100,000$ events from a galactic SN \cite{HyperK}.   Super-K and Hyper-K will measure very well the electron anti-neutrino $\bar\nu_e$ flux and energy spectrum.

The Deep Underground Neutrino Experiment (DUNE) involves a 40 kiloton liquid Ar detector that is planned for the Homestake gold mine in South Dakota.  DUNE should be excellent at detecting electron {\bf neutrinos} $\nu_e$ via $\nu_e+^{40}Ar\rightarrow  ^{40}K^* + e$, and hence should do a good job measuring the flux and spectrum of $\nu_e$.  DUNE ($\nu_e$) and Super-K / Hyper-K ($\bar\nu_e$) provide complimentary information for both neutrinos and anti-neutrinos that is important for nucleosynthesis.

\subsection{Supernovae and the site of the r-process} 
According to many textbooks, r-process nucleosynthesis of the heaviest elements occurs in SN.  Why are the textbooks likely wrong?  The answer may involve some simple neutrino physics \cite{Horowitz:2001yv}.  An important nucleosynthesis site is the neutrino driven wind in a SN where the intense neutrino flux blows some baryons off of the protoneutron star.  The composition of this wind (ratio of neutrons to protons) is set by the relative rates of neutrino and antineutrino captures: $\nu_e+n\rightarrow p + e$ versus $\bar\nu_e+p\rightarrow n + e^+$.   These capture cross sections grow with energy so the relative rates are very sensitive to the relative energies of $\nu_e$ compared to $\bar\nu_e$.  If the $\nu_e$ have considerably lower energies than the $\bar\nu_e$ the wind can be signifcantly neutron rich.  This could allow the production of heavy r-process nuclei.  However modern SN simulations find that the $\nu_e$ are only slightly lower in energy than the $\bar\nu_e$.  As a result the wind is not neutron rich enough to produce heavy r-process nuclei.  Furthermore this conclusion seems to be robust to reasonable changes in the neutrino physics and to details of the SN simulation.  

\begin{figure}[ht]
\smallskip
\includegraphics[width=1.\columnwidth]{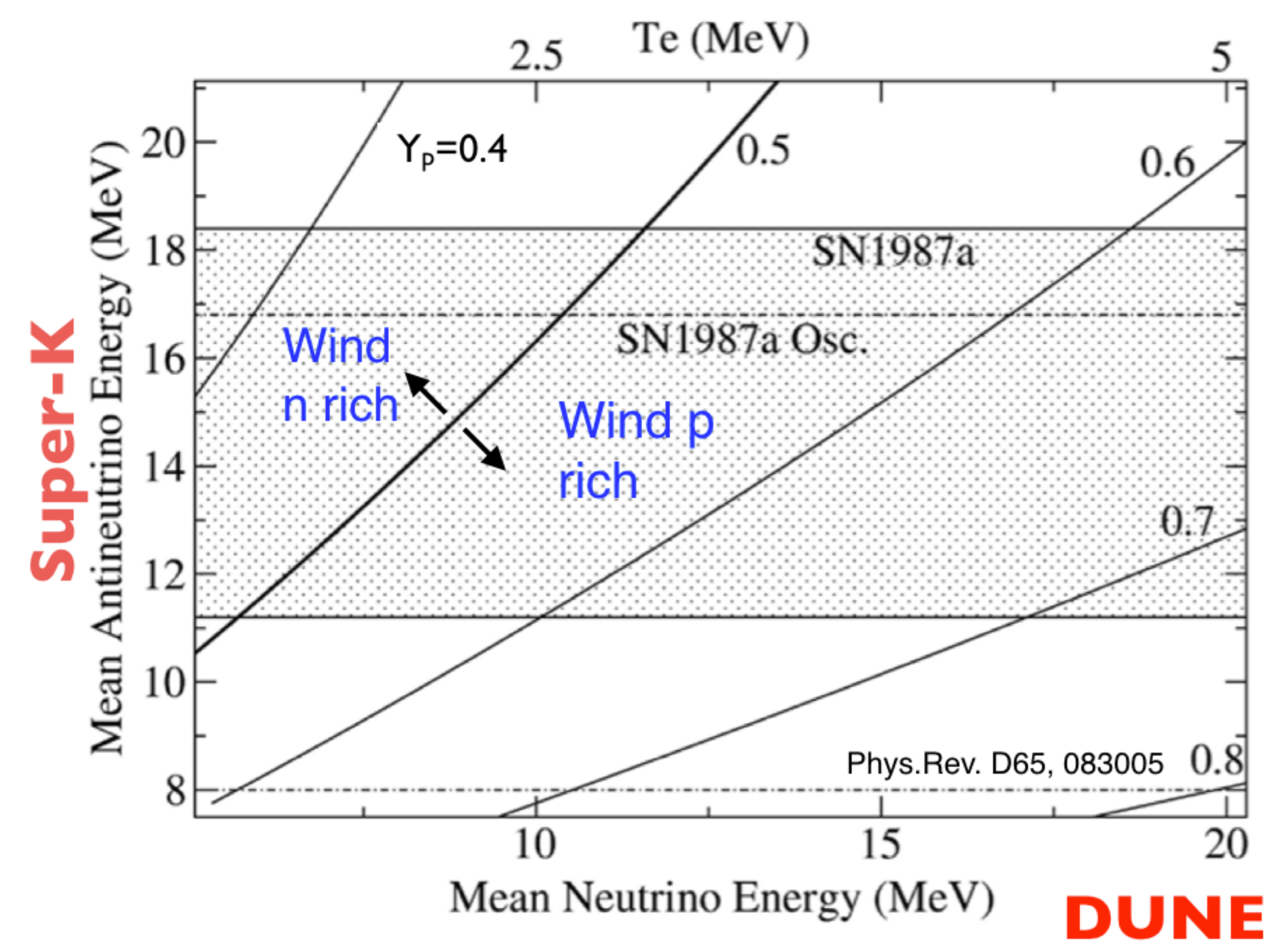}
 \caption{(Color online) 
Contours of proton fraction $Y_p$ (which is equal to electron fraction $Y_e$) of the neutrino driven wind in a SN versus the mean $\nu_e$ energy (X-axis) and $\bar\nu_e$ energy (Y-axis).  The shaded region shows approximate information on mean $\bar\nu_e$ energy from SN1987A.  Figure from ref. \cite{Horowitz:2001yv}}.
\label{Fig1}
\end{figure}
%%%%%%%%%%%%%%

Perhaps there could be some surprise such as an unexpected neutrino oscillation that could change this conclusion and make the wind neutron rich enough for the heavy r-process.   Therefore, it is very important, for the next galactic SN, to measure separately the $\nu_e$ spectrum (as can be done with DUNE) and the $\bar\nu_e$ spectrum from Super-K or Hyper-K.  This information can help determine the composition of the wind and the expected nucleosynthesis.  This is illustrated in Fig.~\ref{Fig1}.      

\subsection{The r-process after GW170817}

We conclude that the r-process does not occur in SN because simple neutrino physics keeps the ejecta from being very neutron rich.  Instead, the red spectrum at late times of the kilonova associated with GW170817 strongly suggests that NS mergers are an important site for the main (heavy) r-process.  This is fundamental progress but there are still many open questions.  First, what are the amounts, velocities, and compositions ($Y_e$) of the different ejecta components during a NS merger.  These components can include ejecta from tidal tails, collisions, neutrino driven winds, accretion disk evaporation ...

Second, what chemical elements, and in what abundances, were actually produced?  The very red color suggests that lanthanides were produced.  But which ones? Perhaps we may be able to identify particular elements in future kilonova observations using spectra from very large telescopes.  However, this may be difficult because of high ejecta velocities and large doppler broadening of any spectral lines.          
 
Finally, do calculations of nucleosynthesis in NS merger ejecta actually reproduce the solar system r-process abundances?  At the moment there are important nuclear physics unknowns that greatly increase the uncertainties in abundance predictions.  The unknown nuclear physics includes the masses, beta decay half-lives, and neutron capture cross sections of very neutron rich heavy nuclei.  Radioactive beam accelerators such as the Facility for Rare Isotope Beams (FRIB) can produce and study these nuclei.  Indeed as discussed in a National Academy study, FRIB was funded in part because it can produce many of the exotic nuclei involved in the r-process.  LIGO, on the other hand, was funded in part to observe gravitational waves from NS mergers.  It now appears that these mergers are an important r-process site.  Therefore both FRIB and LIGO were funded in part to study the same events, in extraordinarily different ways.  This scientific overlap provides an important opportunity for both FRIB and LIGO.  An extensive review article on the r-process discusses many of these issues \cite{Horowitz:2018ndv}.

\section{Equation of state}
\label{Sec3}
The equation of state (EOS), or pressure as a function of density, of neutron rich matter is important for modeling NS and NS mergers.  At low densities, nuclear structure observables and in particular the neutron skin thickness of a heavy nucleus probes the EOS near normal nuclear density $\rho_0=3\times 10^{14}$ g/cm$^3$.  The radius $R_{ns}$ or the gravitational deformability $\Lambda$ of a NS probes the EOS at medium densities near $2\rho_0$.  Finally the maximum mass of a NS, before the system collapses to a black hole, is sensitive to the EOS at high densities $>2\rho_0$.  Taken together these three kinds of observables can map out the density dependence of the EOS.

A neutron star is 18 orders of magnitude larger than a Pb nucleus but both systems contain the same neutrons, with the same strong interactions, and are governed by the same EOS.  Therefore a measurement in one domain, be it astrophysics or nuclear physics, can have important implications in the other domain.   A laboratory observable that has been identified as strongly correlated to
both the deformability $\Lambda$ and the radius $R_{ns}$ of neutron stars is the
\emph{neutron-skin thickness} of atomic nuclei---defined as the difference
between the neutron ($R_{n}$) and proton ($R_{p}$) root-mean-square radii:
$R_{\rm skin}\!=\!R_{n}\!-\!R_{p}$. Despite a difference in length scales of
18 orders of magnitude, the size of a neutron star and the thickness
of the neutron skin share a common origin: the pressure of
neutron-rich matter. That is, whether pushing against surface
tension in an atomic nucleus or against gravity in a neutron star,
both the neutron skin and the stellar radius are sensitive to the
same EOS.

 \begin{figure}[ht]
\smallskip
\includegraphics[width=0.8\columnwidth]{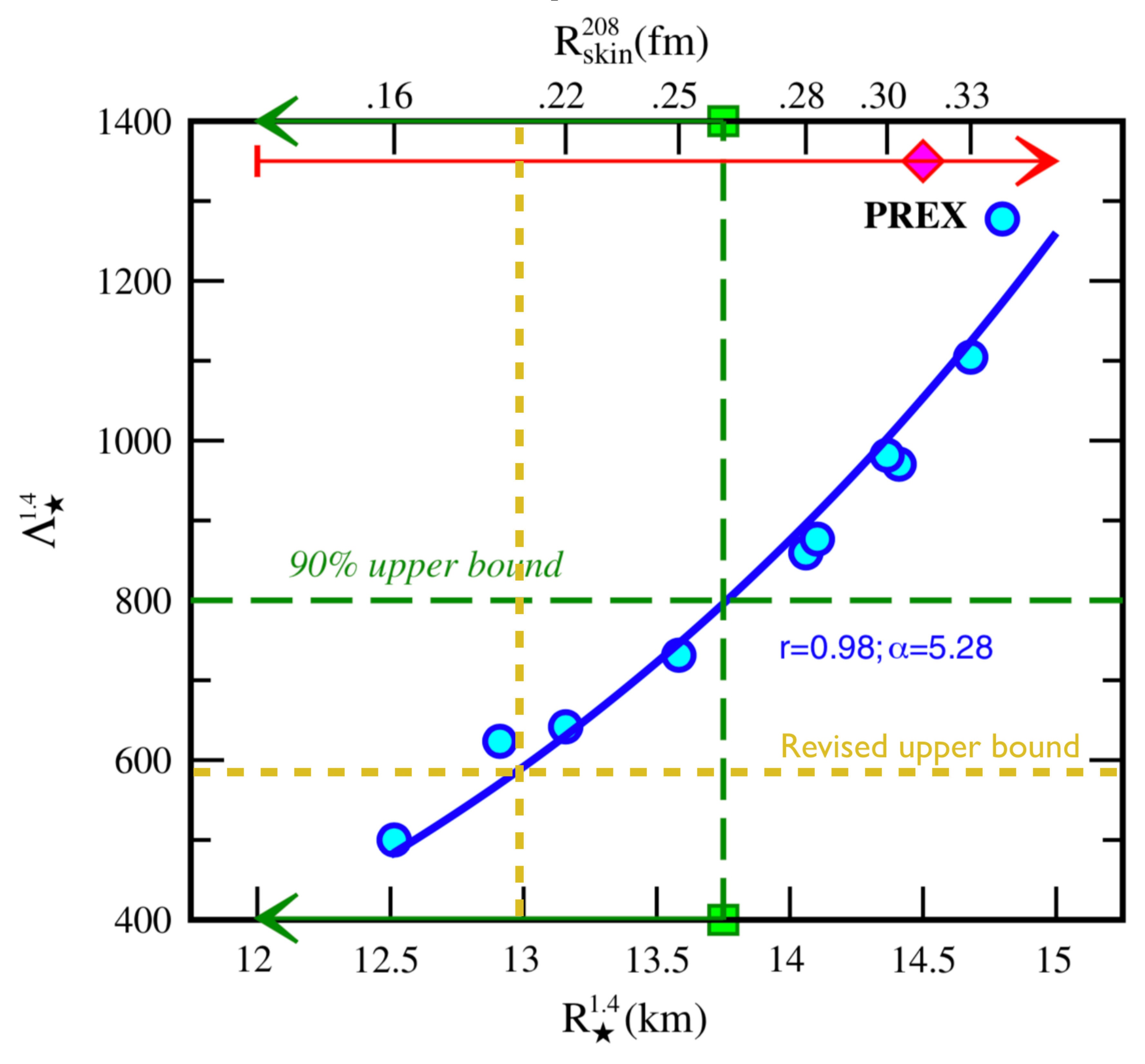}
 \caption{(Color online) 
Gravitational deformability $\Lambda_*^{1.4}$ is shown on the Y-axis versus the radius $R_*^{1.4}$ of a 1.4$M_\odot$ NS on the lower X-axis.  Also shown is the neutron skin in ${}^{208}$Pb on the upper X-axis.  The blue circles are results for a series of model relativistic energy density functionals.  The green dashed line shows the original upper bound on $\Lambda$ from GW170817 data.  Finally the dotted yellow line shows a possible more stringent upper bound on $\Lambda$ from the GW170817 data and assuming the two NS have the same EOS. Figure adopted from ref. \cite{PhysRevLett.120.172702}.}
\label{Fig2}
\end{figure}
%%%%%%%%%%%%%%

The pioneering Lead Radius Experiment (PREX) at the
Jefferson Laboratory has provided the first model-independent evidence
in favor of a neutron-rich skin in
${}^{208}$Pb\,\cite{PhysRevLett.108.112502,PhysRevC.85.032501}:
%$R_{\rm skin}^{208}\!=\!{0.33}^{+0.16}_{-0.18}\,{\rm fm}.$
%%%
\begin{equation}
   R_{\rm skin}^{208}\!\equiv\!
   R_{n}({}^{208}{\rm Pb}) - R_{p}({}^{208}{\rm Pb}) =
   {0.33}^{+0.16}_{-0.18}\,{\rm fm}.
  \label{PREX}
\end{equation}
%%%
Although the central value is significantly larger than suggested by
most theoretical predictions, the large statistically-dominated
uncertainty prevents any real tension between theory and experiment.
In an effort to impose meaningful theoretical constraints, an approved
follow-up experiment (PREX-II) is envisioned to reach a $0.06$\,fm
sensitivity.  PREX-II is now scheduled to run in the summer of 2019 at Jefferson Laboratory.

Figure~\ref{Fig2} shows expectations for the gravitational deformability $\Lambda_*^{1.4}$ and the radius $R_*^{1.4}$ of a 1.4$M_\odot$ NS as calculated for a series of model relativistic energy density functionals (blue circles) \cite{PhysRevLett.120.172702}.  Also shown are expectations for the neutron skin in $R_{\rm skin}^{208}$ of ${}^{208}$Pb on the upper X-axis.  We see that an upper bound on $\Lambda$ from the GW170817 data imply an upper bound on $R$ and an upper bound on $R_{\rm skin}^{208}$.   The GW170817 data for $\Lambda$ disfavor the large central PREX value for  $R_{\rm skin}^{208}$ of $\approx 0.33$ fm.  However the PREX error bar still allows a region where $R_{\rm skin}^{208}$ is about 0.15 to 0.21 or 0.25 fm that is consistent with both PREX and GW170817.   This will be directly tested with PREX II.   

\begin{figure}[ht]
\smallskip
\includegraphics[width=0.85\columnwidth]{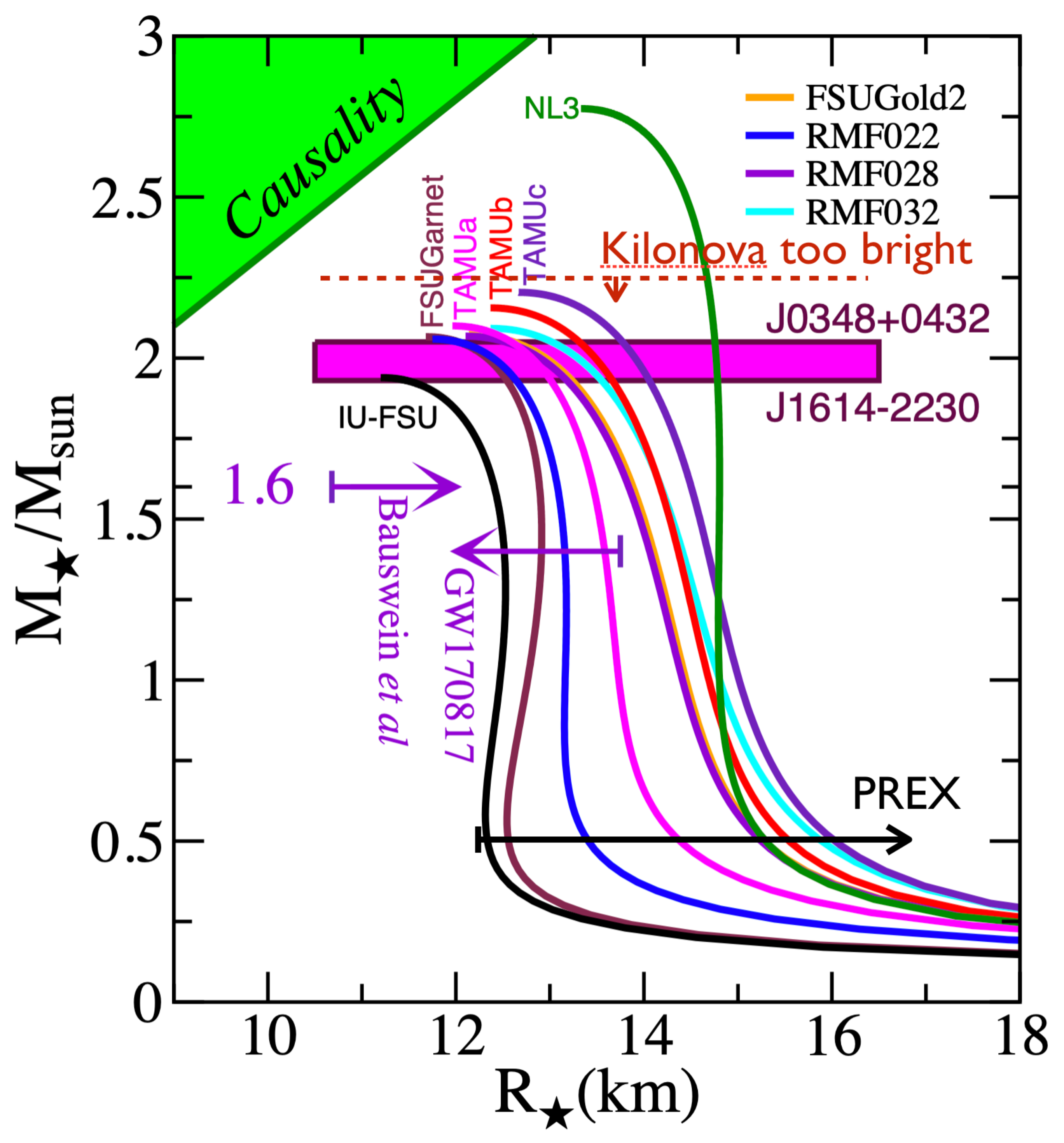}
 \caption{(Color online) 
Neutron star mass versus radius.  The curves are for a variety of model EOS.  Bauswein et al. have suggested if $1.6M_\odot$ stars have radii less than the indicated lower limit, the NS in GW170817 would have collapsed to soon to a black hole and not eject enough material to power the observed kilonova.   If the maximum mass of a NS is above the red dotted line then Metzger et al have argued the compact remnant in GW170817 will live too long and provide too much E+M energy to the kilonova.  Finally the GW170817 limit on radius of a 1.4$M_\odot$ star is from the limit on $\Lambda$ while the limit on the EOS at low density from PREX is plotted as a minimum radius of a 0.5$M_\odot$ NS.  Such low mass NS have low central densities comparable to nuclear density.  Figure adopted from ref. \cite{PhysRevLett.120.172702}.}
\label{Fig3}
\end{figure}
%%%%%%%%%%%%%%

%Note that Fig.~\ref{Fig2} assumes an approximate density dependence to the EOS.  
We caution that the relation between $\Lambda_*^{1.4}$ and $R_{\rm skin}^{208}$ in Fig.~\ref{Fig2} is based on mean field models that assume a simple density dependence for the EOS.  If the EOS has strong density dependence, for example from a phase transition, then this relation could change.  The neutron skin is probing the EOS at low density while $\Lambda_*^{1.4}$ is sensitive to the EOS at higher densities.  If future GW observations confirm that $\Lambda_*^{1.4}$ is small while PREX-II were to confirm that $R_{\rm skin}^{208}$ is large (say near 0.3 fm) then this could indicate a strong density dependence of the EOS where the pressure is relatively large at low densities and then does not increase very much with increasing density so that $\Lambda_*^{1.4}$ remains small.  This reduction in the pressure at high densities could signal a phase transition to, for example, a form of quark matter.  Finally, we collect in Fig.~\ref{Fig3} a number of limits on the mass and radius of NS related to GW170817.

\subsection{The equation of state after GW170817}

The fate of the remnant after a NS merger, and in general how long the remnant takes to collapse to a black hole, is important for the nucleosynthesis, kilonova, and possible gamma ray burst.  In addition, this fate likely depends very sensitively on the combined mass of the two merging NS and how this compares to the maximum mass of a NS.  Mergers with even slightly different NS masses could have very different kilonovae.  The third LIGO / VIRGO observing run is to start soon.  Furthermore as detector sensitivity improves, we expect to observe more (perhaps many more) NS mergers in the near future.  Observing the diversity of kilonova for different NS masses should help constrain the maximum mass of a NS.  In addition, these merger events will provide more deformability information.  Finally, PREX II should accurately measure the neutron skin thickness.  This will provide low density EOS information from the neutron skin, medium density information from the deformability, and high density information from the NS maximum mass.  Together this will allow us to map out the density dependance of the EOS and probe for any possible phase transitions.

\section{Cold dense QCD (quark and gluon) matter}
\label{Sec4}
What are neutron stars made of?  What are the dynamical degrees of freedom of dense neutron rich matter?  Is dense matter best described with hadrons or quarks?  Do strange quarks or hyperons appear?  These important questions go beyond the information contained in the equation of state.  For example if the EOS shows that the pressure is high at high density, this could be because the system contains strongly interacting nucleons or strongly interacting quarks.  In either case the strong interactions would give rise to the high pressure.

Transport properties such as the thermal conductivity or neutrino emissivity and the related heat capacity provide additional information not contained in the EOS and can help constrain the degrees of freedom of dense matter.  We have made the first measurements of the heat capacity of a NS by looking at crust cooling for transiently accreting NS \cite{PhysRevC.95.025806}.   A NS in a low mass X-ray binary (LMXB) may accrete material from its companion for years.  This accretion both heats the star and provides bright X-ray emission so that the total amount of accretion and hence the total heat transferred to the core $\Delta Q$ of the star can be monitored.   Then when accretion stops and the crust cools back into steady state equilibrium with the core, one can infer the final core temperature $T_f$ from X-ray observations of the final surface temperature.   Simple thermodynamics says the heat capacity of the NS core $C$ is equal to the transferred heat $\Delta Q$ over the change in core temperature $\Delta T$,
\begin{equation}
C=\frac{\Delta Q}{\Delta T} > \frac{\Delta Q}{T_f}\, .
\end{equation}   
Often we do not know the original core temperature.  However the final temperature $T_f$ provides an upper limit on $\Delta T$ since the initial temperature is $>0$.  This provides a lower limit on the heat capacity.% of a NS (which is often dominated by the heat capacity of the star's nearly isothermal core).

\begin{figure}[ht]
\smallskip
\includegraphics[width=0.85\columnwidth]{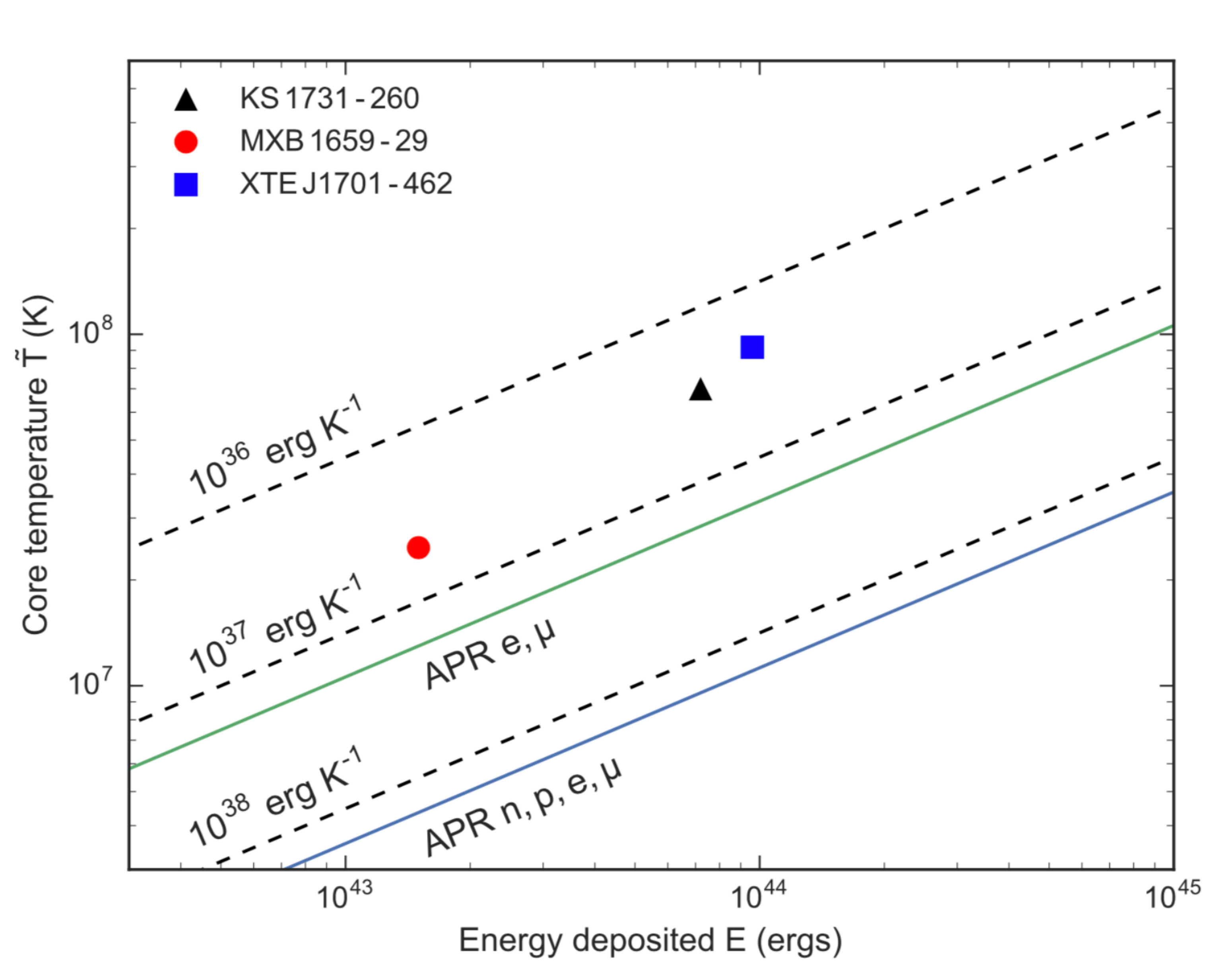}
 \caption{(Color online) 
Core temperature versus energy deposited for accretion episodes on three LMXB (symbols).   The lower blue solid line shows the heat capacity of a NS modeled with the APR EOS assuming all of the nucleons are unpaired.  The green solid line is the heat capacity assuming all of the nucleons are paired.  In this case only electrons and muons contributed to the heat capacity.  The dashed lines show heat capacities of $10^{36}$, $10^{37}$, and $10^{38}$ erg/K.  Figure adopted from ref. \cite{PhysRevC.95.025806}.}
\label{Fig4}
\end{figure}
%%%%%%%%%%%%%%

The lower limits for heat capacities $C$ of three LMXBs are shown in Fig.~\ref{Fig4}.  The heat capacity of a NS modeled with the APR EOS is also shown.  This has contributions from protons, neutrons, electrons and muons.  If all of the nucleons are unpaired the heat capacity is large and shown by the blue curve with $C>10^{38}$ erg/K.  In reality we expect many of the neutrons and protons to be paired into superfluid or superconducting regions where the nucleons will not contribute to $C$ for temperatures less than the critical temperature.  If all of the nucleons are paired then $C$ is given by the green line with only electron and muon contributions.  We expect some of the nucleons to be paired and some unpaired so the heat capacity should be between the green and blue lines.  This corresponds to a heat capacity that is safely larger than the measured limits.  Thus NS composed of neutrons, protons, electrons and muons have heat capacities which are consistent with our observed limits.    

Alternatively, the very high density limit of QCD is expected to be a color superconductor.  This phase as we describe below has a low heat capacity and our observations rule out the most extreme scenario where NS are made of color flavor locked color superconducting matter with a low transition density \cite{PhysRevC.95.025806}.  At very high densities the Fermi momentum is large and QCD interactions are weak at very large momenta.  Therefore matter at extremely high densities is expected to approach a Fermi gas of up, down and strange quarks.  One gluon exchange between these quarks is attractive for some color states and should pair these quarks into a color superconductor.  Because the negative charge is carried on both down and strange quarks this matter can be electrically neutral without any electrons or muons.  With no leptons and all of the quarks paired this color superconductor is expected to have a very low heat capacity.  Indeed in the most extreme case, the heat capacity is below our bound.  Therefore, our observations rule out the most extreme color superconductor being present in NS.   

Neutron star cooling provides additional information on the degrees of freedom of very dense matter.  This is because NS cool by neutrino emission from their dense interiors.  One can observe isolated NS and how they cool after they were born hot in SN.  In addition, one can observe how accreting NS cool after accretion.  We recently observed an accreting NS in the system MXB 1659 that has a very low temperature despite large accretion.  This is the first star, with a well measured surface temperature, that needs rapid neutrino cooling \cite{PhysRevLett.120.182701}.  This suggests that the interior of the star either has a large proton fraction so that neutrons can beta decay quickly or there are additional hadrons beyond protons and neutrons present in the core of the star.  These extra hadrons, such as quarks or hyperons, could then beta decay and radiate neutrinos to cool the star.

\section{Conclusion: Dense matter after GW170817}

The historic observations of the neutron star merger GW170817 advanced our understanding of r-process nucleosynthesis and the equation of state (EOS) of neutron rich matter.  But much remains to be done.  For the r-process we need to measure the masses and beta decay half-lives of very neutron rich heavy nuclei so that we can more accurately predict the abundances of heavy elements that are produced.  This can be done with new radioactive beam accelerators such as the Facility for Rare Isotope Beams (FRIB).

The PREX II experiment will measure the neutron skin of ${}^{208}$Pb and help constrain the low density EOS.  Additional GW observations of NS mergers will better determine the NS deformability and the EOS at medium densities.  Finally GW and electromagnetic observations of kilonovae can better constrain the maximum mass of a NS and the high density EOS.  Perhaps most importantly, as the sensitivity of GW detectors improve, we expect to observe many more events.   Gravitational wave astronomy has already provided important advances.  And it is just getting started.  We look forward to many more exciting advances and surprises!

\section*{Acknowledgements}
This work is supported in part by DOE grants DE-FG02-87ER40365 and DE-SC0018083.


\begin{thebibliography}{10}
\expandafter\ifx\csname url\endcsname\relax
  \def\url#1{\texttt{#1}}\fi
\expandafter\ifx\csname urlprefix\endcsname\relax\def\urlprefix{URL }\fi
\expandafter\ifx\csname href\endcsname\relax
  \def\href#1#2{#2} \def\path#1{#1}\fi
\bibitem{Chiral_review}
R Machleidt and F Sammarruca, Phys. Scr. {\bf 91} (2016) 083007.


\bibitem{LIGO1} 
B. P. Abbott et al. (Virgo, LIGO Scientific), Phys. Rev. Lett. {\bf 119} (2017) 161101.

\bibitem{LIGO2}
B.  P.  Abbott,  R.  Abbott,  T.  D.  Abbott,  F.  Acernese,
K. Ackley, C. Adams, T. Adams, P. Addesso, R. X. Ad-
hikari, V. B. Adya, et al. (The LIGO Scientific Collab-
oration and the Virgo Collaboration),  Phys. Rev. Lett. {\bf 121} (2018) 161101.  

\bibitem{SN_osc}
Huaiyu Duan, George M. Fuller, and Yong-Zhong Qian,  Annual Review of Nuclear and Particle Science {\bf 60} (2010) 569.

\bibitem{PhysRevD.70.043006}
M.~L. Costantini, A.~Ianni, F.~Vissani,
  \href{http://link.aps.org/doi/10.1103/PhysRevD.70.043006}{Sn1987a and the
  properties of the neutrino burst}, Phys. Rev. D 70 (2004) 043006.
\newblock \href {http://dx.doi.org/10.1103/PhysRevD.70.043006}
  {\path{doi:10.1103/PhysRevD.70.043006}}.
\newline\urlprefix\url{http://link.aps.org/doi/10.1103/PhysRevD.70.043006}

\bibitem{Scholberg2012}
K.~Scholberg, Supernova neutrino detection, Ann. Rev. Nuclear and Particle
  Science 62 (2012) 81.

\bibitem{HyperK}
K.~A. et~al, Physics potential of a long-baseline neutrino oscillation
  experiment using a j-parc neutrino beam and hyper-kamiokande, Prog. Theor.
  Exp. Phys. 053C02.

\bibitem{Horowitz:2001yv}
C.~J. Horowitz, {Supernova SN1987A bound on neutrino spectra for R-process
  nucleosynthesis}, Phys. Rev. D65 (2002) 083005.
\newblock \href {http://arxiv.org/abs/astro-ph/0108113}
  {\path{arXiv:astro-ph/0108113}}, \href
  {http://dx.doi.org/10.1103/PhysRevD.65.083005}
  {\path{doi:10.1103/PhysRevD.65.083005}}.

\bibitem{Horowitz:2018ndv}
C.~J. Horowitz, et~al., {r-Process Nucleosynthesis: Connecting Rare-Isotope
  Beam Facilities with the Cosmos}, J. Phys. G in press 2019.
\newblock \href {http://arxiv.org/abs/1805.04637} {\path{arXiv:1805.04637}}.

\bibitem{PhysRevLett.120.172702}
F.~J. Fattoyev, J.~Piekarewicz, C.~J. Horowitz,
  \href{https://link.aps.org/doi/10.1103/PhysRevLett.120.172702}{Neutron skins
  and neutron stars in the multimessenger era}, Phys. Rev. Lett. 120 (2018)
  172702.
\newblock \href {http://dx.doi.org/10.1103/PhysRevLett.120.172702}
  {\path{doi:10.1103/PhysRevLett.120.172702}}.
\newline\urlprefix\url{https://link.aps.org/doi/10.1103/PhysRevLett.120.172702}

\bibitem{PhysRevLett.108.112502}
S.~Abrahamyan, Z.~Ahmed, H.~Albataineh, K.~Aniol, D.~S. Armstrong,
  W.~Armstrong, T.~Averett, B.~Babineau, A.~Barbieri, V.~Bellini,
  R.~Beminiwattha, J.~Benesch, F.~Benmokhtar, T.~Bielarski, W.~Boeglin,
  A.~Camsonne, M.~Canan, P.~Carter, G.~D. Cates, C.~Chen, J.-P. Chen, O.~Hen,
  F.~Cusanno, M.~M. Dalton, R.~De~Leo, K.~de~Jager, W.~Deconinck, P.~Decowski,
  X.~Deng, A.~Deur, D.~Dutta, A.~Etile, D.~Flay, G.~B. Franklin, M.~Friend,
  S.~Frullani, E.~Fuchey, F.~Garibaldi, E.~Gasser, R.~Gilman, A.~Giusa,
  A.~Glamazdin, J.~Gomez, J.~Grames, C.~Gu, O.~Hansen, J.~Hansknecht, D.~W.
  Higinbotham, R.~S. Holmes, T.~Holmstrom, C.~J. Horowitz, J.~Hoskins,
  J.~Huang, C.~E. Hyde, F.~Itard, C.-M. Jen, E.~Jensen, G.~Jin, S.~Johnston,
  A.~Kelleher, K.~Kliakhandler, P.~M. King, S.~Kowalski, K.~S. Kumar,
  J.~Leacock, J.~Leckey, J.~H. Lee, J.~J. LeRose, R.~Lindgren, N.~Liyanage,
  N.~Lubinsky, J.~Mammei, F.~Mammoliti, D.~J. Margaziotis, P.~Markowitz,
  A.~McCreary, D.~McNulty, L.~Mercado, Z.-E. Meziani, R.~W. Michaels,
  M.~Mihovilovic, N.~Muangma, C.~Mu\~noz Camacho, S.~Nanda, V.~Nelyubin,
  N.~Nuruzzaman, Y.~Oh, A.~Palmer, D.~Parno, K.~D. Paschke, S.~K. Phillips,
  B.~Poelker, R.~Pomatsalyuk, M.~Posik, A.~J.~R. Puckett, B.~Quinn, A.~Rakhman,
  P.~E. Reimer, S.~Riordan, P.~Rogan, G.~Ron, G.~Russo, K.~Saenboonruang,
  A.~Saha, B.~Sawatzky, A.~Shahinyan, R.~Silwal, S.~Sirca, K.~Slifer,
  P.~Solvignon, P.~A. Souder, M.~L. Sperduto, R.~Subedi, R.~Suleiman,
  V.~Sulkosky, C.~M. Sutera, W.~A. Tobias, W.~Troth, G.~M. Urciuoli,
  B.~Waidyawansa, D.~Wang, J.~Wexler, R.~Wilson, B.~Wojtsekhowski, X.~Yan,
  H.~Yao, Y.~Ye, Z.~Ye, V.~Yim, L.~Zana, X.~Zhan, J.~Zhang, Y.~Zhang, X.~Zheng,
  P.~Zhu,
  \href{https://link.aps.org/doi/10.1103/PhysRevLett.108.112502}{Measurement of
  the neutron radius of $^{208}\mathrm{Pb}$ through parity violation in
  electron scattering}, Phys. Rev. Lett. 108 (2012) 112502.
\newblock \href {http://dx.doi.org/10.1103/PhysRevLett.108.112502}
  {\path{doi:10.1103/PhysRevLett.108.112502}}.
\newline\urlprefix\url{https://link.aps.org/doi/10.1103/PhysRevLett.108.112502}

\bibitem{PhysRevC.85.032501}
C.~J. Horowitz, Z.~Ahmed, C.-M. Jen, A.~Rakhman, P.~A. Souder, M.~M. Dalton,
  N.~Liyanage, K.~D. Paschke, K.~Saenboonruang, R.~Silwal, G.~B. Franklin,
  M.~Friend, B.~Quinn, K.~S. Kumar, D.~McNulty, L.~Mercado, S.~Riordan,
  J.~Wexler, R.~W. Michaels, G.~M. Urciuoli,
  \href{https://link.aps.org/doi/10.1103/PhysRevC.85.032501}{Weak charge form
  factor and radius of ${}^{208}$pb through parity violation in electron
  scattering}, Phys. Rev. C 85 (2012) 032501.
\newblock \href {http://dx.doi.org/10.1103/PhysRevC.85.032501}
  {\path{doi:10.1103/PhysRevC.85.032501}}.
\newline\urlprefix\url{https://link.aps.org/doi/10.1103/PhysRevC.85.032501}

\bibitem{PhysRevC.95.025806}
A.~Cumming, E.~F. Brown, F.~J. Fattoyev, C.~J. Horowitz, D.~Page, S.~Reddy,
  \href{https://link.aps.org/doi/10.1103/PhysRevC.95.025806}{Lower limit on the
  heat capacity of the neutron star core}, Phys. Rev. C 95 (2017) 025806.
\newblock \href {http://dx.doi.org/10.1103/PhysRevC.95.025806}
  {\path{doi:10.1103/PhysRevC.95.025806}}.
\newline\urlprefix\url{https://link.aps.org/doi/10.1103/PhysRevC.95.025806}

\bibitem{PhysRevLett.120.182701}
E.~F. Brown, A.~Cumming, F.~J. Fattoyev, C.~J. Horowitz, D.~Page, S.~Reddy,
  \href{https://link.aps.org/doi/10.1103/PhysRevLett.120.182701}{Rapid neutrino
  cooling in the neutron star mxb 1659-29}, Phys. Rev. Lett. 120 (2018) 182701.
\newblock \href {http://dx.doi.org/10.1103/PhysRevLett.120.182701}
  {\path{doi:10.1103/PhysRevLett.120.182701}}.
\newline\urlprefix\url{https://link.aps.org/doi/10.1103/PhysRevLett.120.182701}

\end{thebibliography}
\end{document}